\documentclass{ar-1col-S2O}
\usepackage[numbers]{natbib}
\usepackage{url}
\usepackage[english]{babel}
\usepackage[version=4]{mhchem}
\usepackage{physics}
\usepackage{textgreek}
\usepackage{silence}
\usepackage{enumitem}

\newcommand\ri{\right}
\renewcommand\le{\left}

\renewcommand\a{\alpha}
\renewcommand\b{\beta}
\renewcommand\k{\kappa}

\jname{Xxxx. Xxx. Xxx. Xxx.}
\jvol{AA}
\jyear{YYYY}
\doi{10.1146/((please add article doi))}

\begin{document}

\markboth{Wright, Joshi, et al.}{Emergent Simplicities}

\title{Emergent Simplicities in the Living Histories of Individual Cells}

\author{Charles S. Wright,$^{1,2,\ast}$ Kunaal Joshi,$^{1,\ast}$ Rudro R. Biswas,$^{1,\dagger}$ and Srividya Iyer-Biswas$^{1,\dagger}$
\affil{$^1$Department of Physics and Astronomy, Purdue University, West Lafayette, IN 47907, USA}
\affil{$^2$Monash Biomedicine Discovery Institute, Faculty of Medicine, Nursing and Health Sciences, Monash University, Clayton/Melbourne, VIC 3800, Australia}
\affil{$^\ast$These authors contributed equally to this work.} 
\affil{$^\dagger$To whom correspondence should be addressed: rrbiswas@purdue.edu and iyerbiswas@purdue.edu.}
}

\WarningFilter{xcolor}{Incompatible color definition}
\WarningFilter{latex}{Overful}
\WarningFilter{latex}{Underful}

\begin{abstract}
Organisms maintain the status quo, holding key physiological variables constant to within an acceptable tolerance, and yet adapt with precision and plasticity to dynamic changes in externalities. What organizational principles ensure such exquisite yet robust control of systems-level ``state variables'' in complex systems with an extraordinary number of moving parts and fluctuating variables? Here we focus on these issues in the specific context of intra- and intergenerational life histories of individual bacterial cells, whose biographies are precisely charted via high-precision dynamic experiments using the SChemostat technology. We highlight intra- and intergenerational scaling laws and other ``emergent simplicities'' revealed by these high-precision data. In turn, these facilitate a principled route to dimensional reduction of the problem, and serve as essential building blocks for phenomenological and mechanistic theory. Parameter-free data-theory matches for multiple organisms validate theory frameworks, and explicate the systems physics of stochastic homeostasis and adaptation.
\end{abstract}

\maketitle

\tableofcontents

\section{INTRODUCTION}
\label{sec:intro}

If the fundamental purpose of life is to be fruitful and multiply, the humble bacterial cell fulfills this injunction with unparalleled zeal, growing and dividing with alacrity across diverse environments. Bacterial growth dynamics are a stochastic process characterized by continuous growth punctuated at semiregular intervals by division of one cell into two. Yet, even though growth and division are inherently stochastic processes with significant fluctuations, key physiological variables such as cell size remain tightly controlled via an exquisite (yet poorly understood) coupling between growth and division. Despite the long and illustrious history of quantitative bacterial growth studies, basic questions remain about the strategies employed by an organism to negotiate and leverage the environment it happens be in, as well as how those strategies are adapted to dynamic environments. For example, under constant conditions, the path by which cells achieve stochastic homeostasis of the most basic state variables (cell size and growth rate) remains a topic of much debate. Here, we present exciting new developments that are shedding light on the system physics of complex and adaptive systems as revealed by stochastic growth and division processes in bacteria. High-precision dynamic experiments now allow researchers to chart the multigenerational biographies of individual cells in precisely tunable environments, revealing various ``emergent simplicities'' (such as scaling laws) that provide a principled route to dimensional reduction of this problem.

Flexible yet robust architecture is a common feature of functional complex and adaptive systems~\cite{Kirschner1998}. Living systems across length- and timescales make use of a modular approach enabled by specific shared constraints that ensure proper functioning of the organism, while also deconstraining (i.e., conferring flexibilities upon) other aspects of the systems design without introducing risk of catastrophic consequences ~\cite{Doyle2011}. This keeps the system functioning, while remaining adaptive and evolvable. Identification of these constraints provides a complementary framework to prevailing bottom-up mechanistic approaches and serves to directly probe core aspects of how the system operates~\cite{1972-anderson}. But in practice, how can we set about identifying these ``constraints that deconstrain''? A promising approach is to design high-throughput, high-precision experiments to directly record the relevant dynamical behaviors, then sift the data for patterns of temporal organization that may provide helpful clues. However, to pursue this strategy, both technological and conceptual challenges must be overcome. To illustrate, we turn to the phenomenology of homeostasis, the self-regulating process wherein a complex system maintains its internal stability by holding key physiological variables constant within a desired tolerance~\cite{Cannon1926}. This is similar to mechanical self-regulating control systems (such as thermostats); indeed, the mathematical framework describing feedback control in each process shares a common origin~\cite{Wiener1948}. However, any characterization of organismal homeostases from a systems design point of view must distinguish between two opposing paradigms: passive ``elastic adaptation'' characterized by reflexive responses to present conditions with no memory of the past (as in Ashby's Homeostat~\cite{Ashby1960}), versus active ``plastic adaptation'' involving reflective responses achieved by controlled memory integrators operating on information stored from past experiences (as in Shannon's ``Maze Solving Machine''~\cite{Shannon1969}). In contrast, most extant theoretical models of homeostasis have approached the problem using a (quasi-)deterministic framework built on the assumption of a defined ``setpoint'' value of the state variable; this approach captures the behaviors of the mythical ``average'' cell, yet cannot generate a quantitative picture of the stochastic homeostasis of a realistic individual cell, and thus falls short of proving mechanistic insights. To understand the origins of this conundrum, and motivate the pathway to move beyond it, we briefly review the history of bacterial growth physiology and the feedback between experimental and theoretical advances.

In the 1940s, novel techniques for continuous culture of bacterial colonies such as the Chemostat~\cite{1950-novick-szilard-chemostat} permitted researchers to maintain populations at a constant rate of growth for indefinite periods of time~\cite{Monod1950}. These studies of asynchronous bacterial populations in steady state---typically referred to as ``balanced growth''~\cite{Campbell1957}---enabled the first reproducible and quantitative growth studies, leading to the seminal discoveries that total population growth is linearly proportional to the initial concentration of the rate-limiting nutrient~\cite{Monod1949} and that average cell size is exponentially proportional to average nutrient-imposed growth rate~\cite{Schaechter1958}. Shortly thereafter, researchers quantified the growth curves of single cells using time-lapse microscopy of small populations of cells growing over 3--4 generations on an agar surface sandwiched inside a sealed chamber~\cite{Schaechter1962}. Technical limitations prevented determination of the precise mode of growth, but observations were consistent with the hypothesis that cells grow exponentially at roughly the same rate as the whole population~\cite{Koch1962}. The next major experimental refinement occurred with the introduction of the ``Baby Machine''~\cite{Helmstetter1964}, a device for creating a population of cells synchronized according to the most recent cell division. This method permitted the first studies into the timing of events relative to the cell cycle, demonstrating that the onset of DNA replication varies with cell growth rate~\cite{Cooper1968} and occurs when the cell mass per chromosome reaches a (constant) critical value, thus explaining the previously observed relationship between average cell size and growth rate~\cite{Donachie1968}.

For decades, however, high-precision single-cell measurements were accessible only through snapshots of cells drawn from steady-state populations after being ``fixed'' (i.e., preserved in a life-like state for imaging). This approach yielded important results, including the observation that mean-rescaled (asynchronous) cell size distributions obtained from various growth conditions undergo a scaling collapse~\cite{Trueba1982}. However, single-cell growth studies capable of obtaining high-precision, dynamic information did not become widespread until the early 2000s, when researchers developed protocols to more readily and reproducibly integrate growth of cells on flat agarose pads with time-lapse microscopy to obtain movies of individual living cells, allowing for the study of intracellular spatiotemporal organization and morphology~\cite{Fiebig2006}, single-cell growth in dynamic environments~\cite{Ducret2009}, and cell size homeostasis~\cite{Campos2014}. Due to its ease of use, this technique remains indispensable for high-throughput screens linking genotype to phenotype, permitting assays across thousands of combinations of bacterial mutants and growth conditions to link molecular-level details to cell size and growth~\cite{govers2024apparent}. However, exponential growth of cell numbers limits the duration and interpretability of such experiments~\cite{PotvinTrottier2018}---especially problematic when growth itself is the process of interest. The introduction of microfluidic technology sought to address this problem by enabling growth of microcolonies in a precisely controlled chemostat-like environment compatible with long-term imaging~\cite{Weibel2007}. A variety of approaches have been explored~\cite{Balaban2004,Robert2010,Moffitt2012}, but the design that has proved the most popular, known as the ``Mother Machine'', cultivates thousands of cell lineages inside microscope channels open to flowing growth medium at one end and closed at the other; although daughter cells eventually exit the channel, trapped mother cells may be observed for over 100 generations~\cite{Wang2010}. This technology has been applied to  studies of single-cell aging (or lack thereof)~\cite{Wang2010,Lapinska2019}, cell-fate decision making~\cite{Norman2013}, and cell size homeostasis~\cite{taheri2015cell,Si2017,Susman2018,Si2019} at steady state, as well as physiological growth~\cite{Basan2020} and adaptation~\cite{Bakshi2021} in time-varying environments. A complementary approach is the SChemostat~\cite{2014-iyer-biswas-dn} design, which uses controllable surface adhesion to generate a stable population of cells of a defined density chosen to eliminate neighbor--neighbor contacts and ensure that all cells experience an identical flow environment. By removing potential confounding factors such as mechanical confinement or variable microenvironments~\cite{Yang2018}, this permits observation of a statistically identical set of non-interacting cells under precisely controlled and tunable environmental conditions for indefinite periods of time, a prerequisite for detailed studies of stochastic growth and division over many generations.

The physics approach to problem solving has previously proved useful in biological contexts involving stochastic dynamics. For instance, the prevalence of stochasticity in gene expression was revealed through high-precision live-cell imaging, and inspired a new generation of enquiry into the basic biology encapsulated in the central dogma of molecular biology~\cite{2002-elowitz-swain,2002-vanOudenaarden,2003-AvgCell}. Parallel theory development in turn provided mechanistic insights, identification of the important dimensionless parameters governing the phenomenology, natural timescales dictating different path behaviors, and unifying principles. In part, these insights were facilitated through development of new conceptual frameworks attuned to uncover patterns and organization implicit in trajectories that may be obscured in snapshots~\cite{2002-IntExtNoise,2004-paulsson,2008-pedraza,2009-iyer-biswas-fj,2014-iyer-biswas-fj,2009-hu-fv,2009-iyer-biswas-qy,2017-IntThresh}. In this review, we describe how a similar integrated quantitative--analytical--theoretical approach has been applied to the study of stochastic growth and division as manifested through the living histories of individual cells, and illustrate the importance of identified emergent simplicities in yielding analytically tractable and novel insights that point toward fundamental constraints on the architecture of living systems. We end by discussing future directions, applicability to other systems, and systems design implications of these findings.

\section{INTRAGENERATIONAL SCALING LAWS GOVERNING STOCHASTIC GROWTH AND DIVISION OF INDIVIDUAL CELLS}\label{sec:part1intra}

In this section, we review intragenerational scaling laws, that is, quantitative relationships that hold within a given ``generation'' of a single cell's lifetime, defined as the interval between successive division events. Cell size at the first (last) timepoint within a specified generation is referred to as size at birth (division) or as initial (final) size. Note also that most bacterial growth studies are conducted using a small number of ``model'' species. \textit{Escherichia coli}, a rod-shaped bacterium that divides symmetrically into equal-sized daughter cells, is by far the most widely used species. Another popular choice is \textit{Bacillus subtilis}, which typically reproduces by symmetric division but may also undergo asymmetric division (sporulation) when exposed to nutritional stress. \textit{Caulobacter crescentus}, a crescent-shaped bacterium that divides asymmetrically into a replication-competent stalked cell and a smaller, motile swarmer cell (which differentiates into a stalked cell under favorable conditions), has emerged as a model for studies of cell-cycle regulation. In the following discussions of \textit{C. crescentus} data obtained using the SChemostat technology, we note that all observations are, by design, collected from a population composed exclusively of stalked cells.

\subsection{Self organization of intragenerational stochastic exponential growth}

\subsubsection{Historical background}
For bacterial cells, the central problem connected with robustness of cell growth under balanced growth conditions could be stated as follows: How should the mass-conferring components be connected via biochemical reactions such that the ratios between the growing numbers of individual components are held steady, enabling uninterrupted balanced continuance of growth through cell division? An elegant rate equation-based \emph{deterministic} solution to this problem was provided by Hinshelwood over seven decades ago in \cite{1952-hinshelwood}. He showed that if the mass-conferring species of the cell form an autocatalytic loop---the Hinshelwood Cycle---then no matter the initial condition, after some adaptive timescale elapses, all species grow exponentially with a shared exponential growth rate, retaining a fixed ratio of species numbers. Once the steady-state fixed ratios are reached, cell division does not significantly alter these ratios, and within each successive generation, the cell grows at the same exponential rate, which is the geometric mean of the individual rates in the autocatalytic loop.

\subsubsection{Experimental evidence} 
But does cell size, which serves as a proxy for cell mass, in fact grow exponentially between division events? Providing an unequivocal answer through direct experimentation has proved challenging; whereas \emph{population} growth under balanced conditions is readily measured to be exponential and explained as the result of cell doubling at a constant rate~\cite{Monod1949}, establishing the temporal law governing cell size growth remained an open problem for decades. The dominant paradigms considered were linear growth---the simplest possible variation with time---and exponential growth---a ``natural'' choice (see the discussion regarding the Hinshelwood cycle above), with suggestive but inconclusive evidence from early single-cell growth studies~\cite{Schaechter1962}. The difficulty in answering this question arises from the necessity of experimentally resolving the difference between a linear and an exponential curve over a $\sim$2-fold change in cell size between the initial and final points of an individual generation of the bacterial growth curve (requiring much better than $4\%$ resolution~\cite{2014-iyer-biswas-dn}). This is difficult owing to the substantial noise present in bacterial growth curves, even though they clearly show a curvature (i.e., are non-linear) to the naked eye~\cite{Schaechter1962,2014-iyer-biswas-dn}. High-precision SChemostat experiments on large numbers of statistically identical, non-interacting cells, consisting of multigenerational trains of growth curves corresponding to each individual cell (Fig.~\ref{fig:SHC}a), which can be further decomposed into growth curves corresponding to each generation of each cell (Fig.~\ref{fig:SHC}b), allowed this distinction to be resolved via statistical averaging; this established beyond reasonable doubt that the pre-divisional \emph{C. crescentus} stalked cell grows exponentially with time under balanced growth conditions~\cite{2014-iyer-biswas-dn}. Based on such measurements of single-cell size as well as cell mass~\cite{Cermak2016}, it is now considered a widely accepted fact that sizes of pre-divisional bacterial cells, on average, increase exponentially under balanced growth conditions.

\subsubsection{Constraining motif: The Stochastic Hinshelwood Cycle}
\begin{figure}[h]
    \centering
    \includegraphics[width=.65\textwidth]{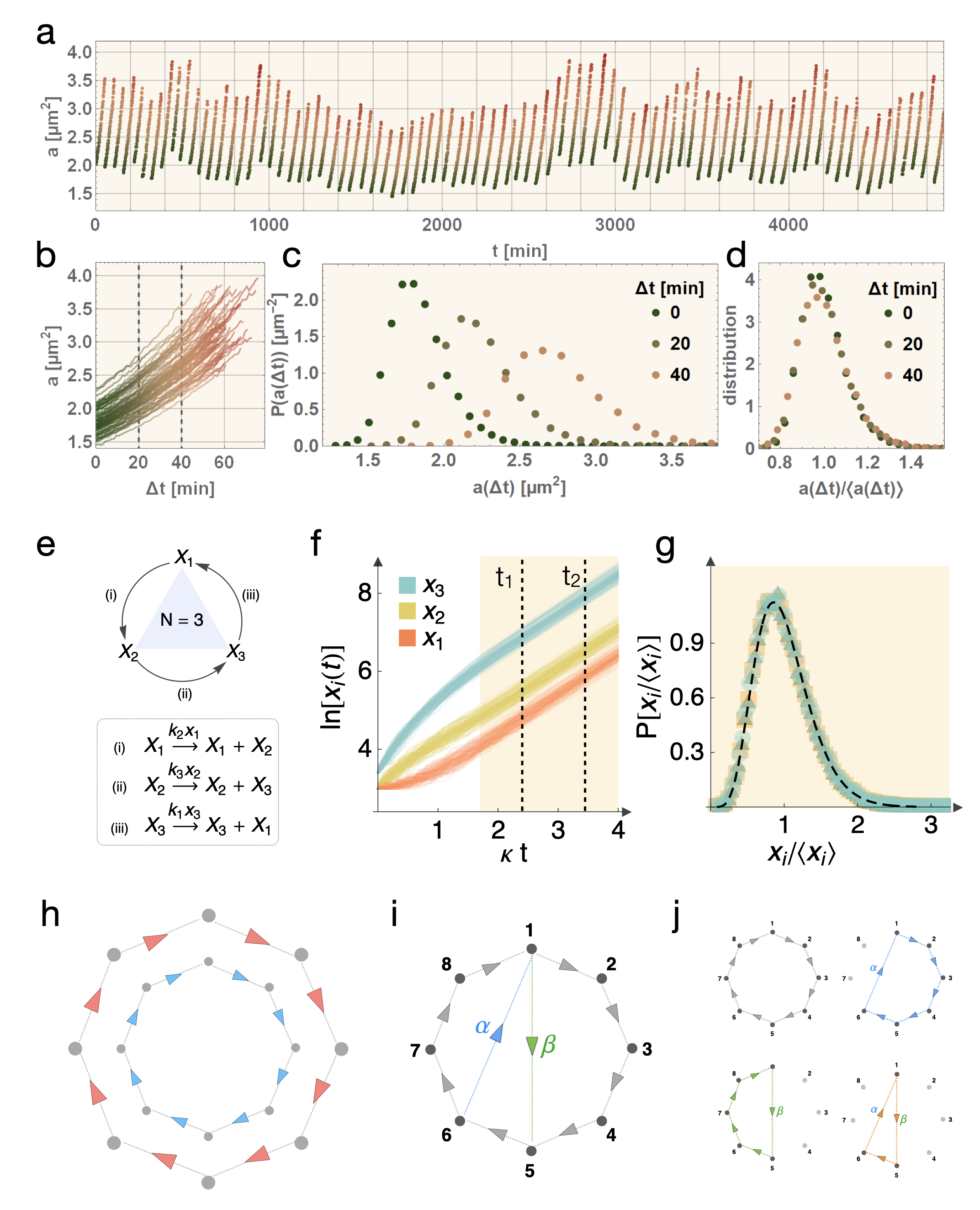}
    \caption{\textbf{Emergent simplicities in stochastic exponential growth of individual cell sizes.} \textbf{(a)} A typical trajectory of cell size (cross-sectional area) for a \textit{C. crescentus} cell in complex media at 34$^\circ$ C, showing successive generations of growth of an individual cell. \textbf{(b)} The same cell's area plotted as a function of cell age (time since last division) for all generations. The growth of cell size within a given generation is exponential. \textbf{(c)} The area distributions at different cell ages are shown, from data pooled from all cells in the experiment. Cell ages correspond to the dashed lines in (b). \textbf{(d)} The distributions in (c), when rescaled by their respective mean values, collapse to the same distribution. This behavior is explained by the Stochastic Hinshelwood Cycle (SHC) model of size growth. \textbf{(e)} A schematic of a three-cycle SHC consisting of an autocatalytic cycle with three constituent species. \textbf{(f)} Gillespie algorithm-simulated trajectories showing the time evolution of the copy numbers of the species in (e). After reaching balanced growth (denoted by yellow background), all species grow exponentially at the same rate (linear on the log-linear scale shown). \textbf{(g)} The mean-rescaled distributions of copy numbers are identical for all species in balanced growth. Total cell size is given by a linear combination of the constituent species' copy numbers, explaining the observed scaling collapse in (d). \textbf{(h)} On increasing the temperature, all rates in the SHC scale by the same factor (given by the Arrhenius scaling law). Hence, the overall growth rate in balanced growth, which is the geometric mean of the individual growth rates, also scales by this factor. The evidence and consequences of this scaling are presented in Fig.~\ref{fig:differenttemperatures} and~\cite{2014-iyer-biswas-dn}. \textbf{(i)} A complex SHC with two extra connections. \textbf{(j)} Loop decomposition of the SHC in (i). The overall growth rate in balanced growth is determined by the loop with the highest geometric mean of individual rates, and all species (including those not in this loop) grow at this rate. Data in (a--e) are taken from~\cite{2014-iyer-biswas-dn}. Data in (e--g) and (i, j) are taken from~\cite{2014-iyer-biswas-qy}.}
    \label{fig:SHC}
\end{figure}

The scaling law governing the intragenerational temporal evolution of cell sizes points to the emergence of an elegant constraining motif from the otherwise confounding complexity and system- and condition-specific details of cellular metabolic networks: the Stochastic Hinshelwood Cycle (SHC)~\cite{2014-iyer-biswas-qy}. First, following the discussion of the deterministic Hinshelwood Cycle above, we see that robustness of intergenerational continuity of exponential growth is satisfied by a noise-free Hinshelwood Cycle scheme of reactions. Inclusion of stochasticity through the most prevalent phenomenological model of stochastic exponential growth, geometric Brownian motion with linear multiplicative noise (which forms the basis of the widely used Black-Scholes equation in economics), predicts runaway noise in cell sizes~\cite{2014-iyer-biswas-qy}. This is manifestly in contradiction with observations. Instead, a generalization of the Hinshelwood cycle that includes stochasticity was found to represent stochastic exponential growth with square-root multiplicative noise~\cite{2014-iyer-biswas-qy}; it belongs to a continuous family of well-behaved models of stochastic exponential growth with fractional power multiplicative noise~\cite{2017-pirjol-qd}. The prediction of square-root multiplicative noise was directly verified from high-precision data~\cite{2014-iyer-biswas-dn}. Another prediction of the SHC model is that mean-rescaled cell sizes follow the same distribution at all times throughout the cell cycle~\cite{2014-iyer-biswas-qy}. As an illustrative example, consider a simplified SHC consisting of three species (Fig.~\ref{fig:SHC}e). Simulations of of the time evolution of the copy numbers of each species demonstrate that all species grow at the same exponential rate at steady state (Fig.~\ref{fig:SHC}f); furthermore, the distributions of copy numbers of each species, upon rescaling by their respective mean values, are identical (Fig.~\ref{fig:SHC}g). This prediction has been validated by a spectacular multidimensional scaling collapse, across generation times and temperatures, of cell sizes from high-precision data~\cite{2014-iyer-biswas-dn}. Thus, the experimentally measured distributions of cell sizes as a function of cell age (defined as time since last division) are distinct for a given cell age (Fig.~\ref{fig:SHC}c); however, after rescaling all distributions by their respective mean values, they collapse onto a single distribution (Fig.~\ref{fig:SHC}d). Yet the question remains: How does the cartoonishly simple Stochastic Hinshelwood Cycle relate to actual complex biochemical networks governing cell growth in system and condition specific ways? We provide the answer in Section~(\ref{sec:SHC2}).

\subsection{Emergence of a cellular unit of time} 
\subsubsection{Implications of the Stochastic Hinshelwood Cycle motif}
Overall population dynamics in a given growth condition are primarily determined by two timescales: that of cell size growth and that of population number growth. The latter is determined by the interdivision time interval (henceforth referred to simply as the ``division time'')~\cite{2018-jafarpour-ad}. Since the SHC predicts that all constituent species grow at the same exponential growth rate in balanced growth, this naturally establishes the timescale of cell size growth as the inverse of this exponential growth rate. This rate is determined by the geometric mean of the reaction rates of the dominant cycle in the SHC. All rates in the SHC are expected to scale by the same factor as temperature varies (provided it remains in the range where balanced growth conditions hold) (Fig.~\ref{fig:SHC}h). Although the connections within the SHC may change between balanced growth conditions (Fig.~\ref{fig:SHC}i), the overall growth rate can be determined by decomposing the network into constitutive loops and identifying the loop with the highest geometric mean of individual rates (Fig.~\ref{fig:SHC}j). Hence, if on varying the ambient temperature the individual rates follow the Arrhenius law with activation energy barriers characteristic of typical enzyme reactions, then so must the composite cell size growth rate! Imposing the constraint of cell size homeostasis couples these two timescales, implying that a single cellular unit of time governs stochastic growth and division of individual cells in a given condition. We address the specifics of this coupling in  Section~(\ref{sec:QX}), but intuitively: if the population numbers grow faster (slower) than the total cell size, the average cell size will decrease (increase).

\subsubsection{Experimental validation of a single cellular unit of time} 
\begin{figure}[h]
    \centering
    \includegraphics[width=.65\textwidth]{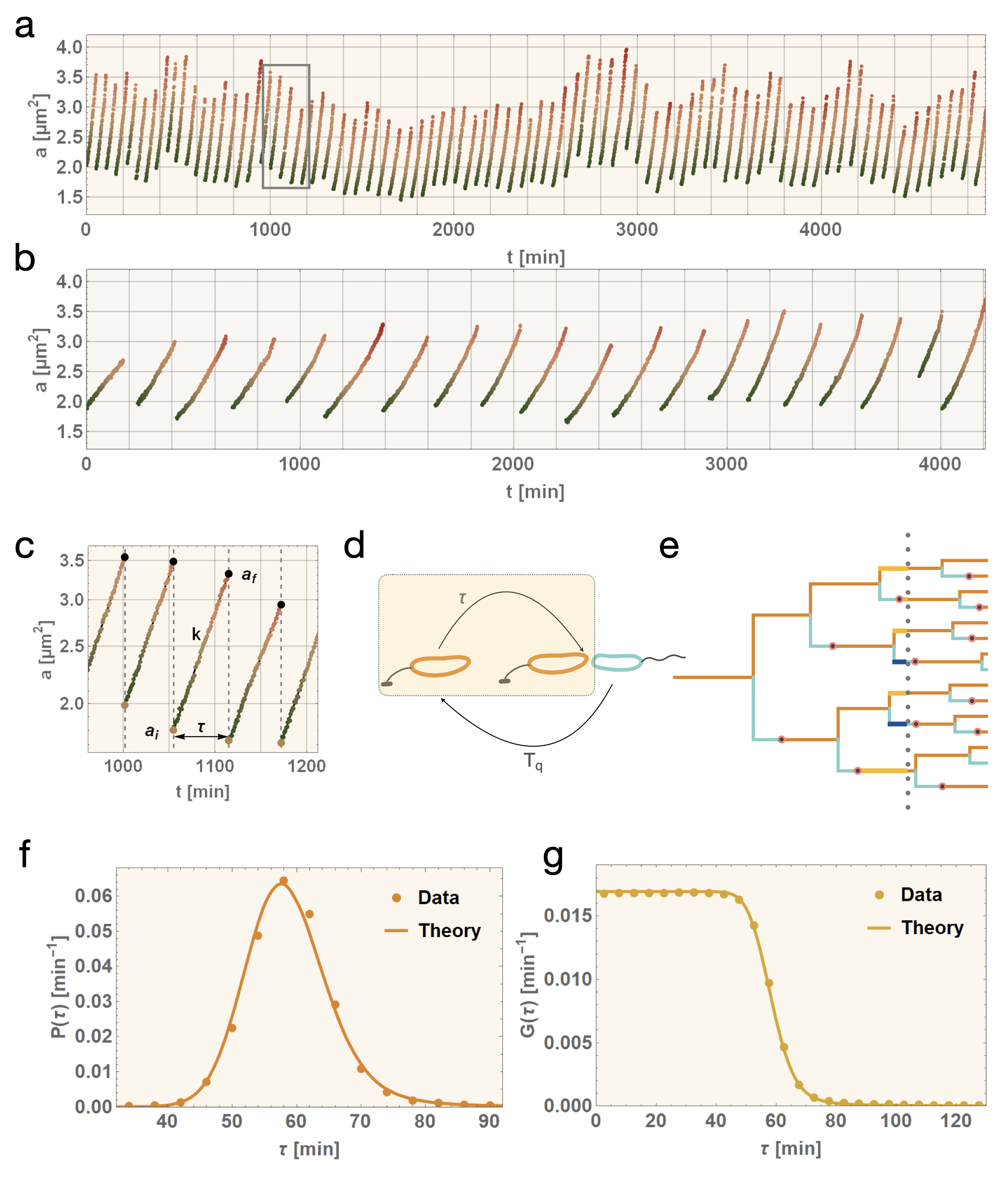}
    \caption{\textbf{An emergent cellular unit of time governs stochastic single-cell growth and division, population growth, and the quiescence timescale in the dimorphic life cycle of \textit{C. crescentus} cells.} \textbf{(a)} A typical cell area trajectory for a \textit{C. crescentus} cell in complex media at 34$^\circ$ C. \textbf{(b)} A typical cell area trajectory for a \textit{C. crescentus} cell in complex media at 17$^\circ$ C. Growth and division are considerably slower than at 34$^\circ$ C, and can be derived from the growth and division dynamics at 34$^\circ$C through an appropriate scaling of a single parameter (see Fig.~\ref{fig:differenttemperatures}). \textbf{(c)} Zoomed inset of the time period highlighted in (a), plotted on a log-linear scale, showing the key parameters: initial size ($a_i$), final size ($a_f$), growth rate ($k$, given by the slope of the linear fit in the log-linear scale), and division time ($\tau$). \textbf{(d)} Schematic showing the timescales relevant to the \textit{C. crescentus} life cycle. The stalked cell divides into a stalked and a swarmer daughter after time $\tau$ after birth. The swarmer cell differentiates into a stalked cell after time $T_q$ after birth. Although both timescales are relevant to population growth, in the single-cell experiments performed using the SChemostat, the swarmer daughter is removed through media flow to prevent crowding; only the cell cycle of the stalked cell is relevant. \textbf{(e)} A tree diagram showing the future progeny lineage of a stalked cell. Stalked cells are represented by orange lines, and swarmer cells by blue lines. The age distribution at a particular time point (dashed line) is given by the times since the last division of all cells present at that time point. \textbf{(f)} The experimentally measured (points) and theoretically predicted (line) steady-state division time distribution for cells in the SChemostat at 34$^\circ$ C. The predictions are based on the model combining the SHC with the thresholding of a protein to trigger constriction. \textbf{(g)} The experimentally measured (points) and theoretically predicted (line) steady-state age distribution for cells in SChemostat at 34$^\circ$ C. Data in (a--c) and (f, g) are taken from~\cite{2014-iyer-biswas-dn} and model predictions from~\cite{2023-architecture}. The schematic in (d) is taken from~\cite{2018-jafarpour-ad}.}
    \label{fig:divisionage}
\end{figure}

The expectation that a single cellular unit of time governs stochastic growth and division of individual cells is spectacularly borne out by experimental data. Cell size trajectories at temperatures spanning the physiologically relevant range for \textit{C. crescentus}, from 34$^\circ$ C (Fig.~\ref{fig:divisionage}a) to 17$^\circ$ C (Fig.~\ref{fig:divisionage}b), show markedly different dynamics. To compare these data, note that each growth curve may be characterized by the initial size ($a_i$), final size ($a_f$), growth rate ($k$), and division time ($\tau$) (Fig.~\ref{fig:divisionage}c). \textit{C. crescentus} has a complex dimorphic life cycle involving asymmetric division into a stalked and a swarmer cell, the latter of which must differentiate into a stalked cell before it is capable of replication (Fig.~\ref{fig:divisionage}d). However, in the SChemostat, swarmer cells are removed immediately after division; thus, out of all possible future lineages of a given cell (Fig.~\ref{fig:divisionage}e), only the single lineage consisting exclusively of stalked cells is observed, and the quiescence timescale need not be considered. For a given condition, the timescale governing growth and division may thus be represented by the distribution of division times (Fig.~\ref{fig:divisionage}f) and the distribution of cell ages (Fig.~\ref{fig:divisionage}g). The coupling of the timescales of size and number growth is evidenced from the observation that the product of the mean growth rate and the mean division time is remarkably invariant across different temperatures~\cite{2014-iyer-biswas-dn}. Furthermore, the mean division time changes with temperature according to the Arrhenius law with an activation energy barrier of approximately $12.9$ kcal/mol~\cite{2014-iyer-biswas-dn} (within the range of temperatures of physiological relevance). This is of the same order of magnitude as the activation energy of a typical enzyme-catalyzed reaction in the cell, providing compelling validation for the SHC model. This Arrhenius-like energy scale of single-cell growth rates matches remarkably well with measurements of population growth rates of multiple species as a function of temperature~\cite{Monod1942,Herendeen1979,Ratkowsky1982}.

\subsubsection{Reconciling the mythical average cell with the stochastic single cell}
The stochastic dynamics of individual cells are not sufficiently represented by simple population averages~\cite{rosenthal2017beyond,sanders2023beyond}. Even among genetically identical cells, the stochasticity inherent in the internal processes and reactions driving cell growth and division causes deviations from the target ideal growth and division dynamics. These deviations, if left unchecked, would result in a breakdown of cellular functions essential for survival; this necessitates control strategies that constrain the propagation of fluctuations and ultimately result in homeostasis. In addition, in physiological contexts, all cells in a given population are not genetically identical, and population heterogeneity plays a significant role in determining control strategies~\cite{lunz2023optimal}. These intricate aspects of cell growth and division dynamics are washed out when simply dealing with population averages. We explore the details of homeostasis of cell size in Section~(\ref{sec:intergenerational}), but now simply note the key requirement for establishing that a single cellular unit of time governs stochastic growth and division of individual cells: the full distributions, and not only the mean values, must scale accordingly.

\begin{figure}[h]
    \centering
    \includegraphics[width=.7\textwidth]{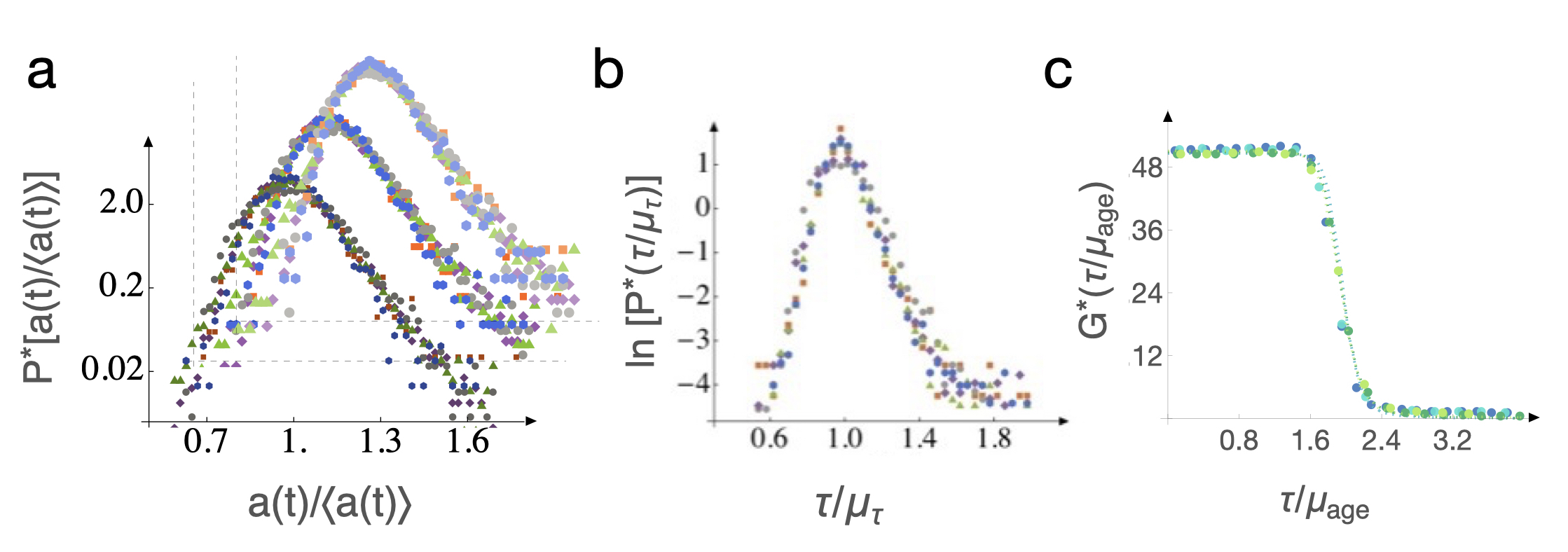}
    \caption{\textbf{Scaling laws governing stochastic growth and division of individual cells on generational timescales.} \textbf{(a)} Mean-rescaled steady-state size (cross-sectional area) distributions for cells at different normalized ages, $t/\langle\tau\rangle=0$, $0.2$, and $0.6$ from back to front. Different colors indicate distributions corresponding to different temperatures (gray, 17$^\circ$ C; blue, 24$^\circ$ C; orange, 28$^\circ$ C; green, 31$^\circ$ C; purple, 34$^\circ$ C). These mean-rescaled distributions are identical irrespective of the age or the temperature, and are shifted for distinguishability. \textbf{(b)} The mean-rescaled division time distributions are identical across growth conditions with different temperatures. The different temperatures follow the same color scheme as in (a). \textbf{(c)} The mean-rescaled age distributions are identical across growth conditions with different temperatures (blue, 17$^\circ$ C; cyan, 24$^\circ$ C; dark green, 31$^\circ$ C; light green, 34$^\circ$ C). These scaling laws show that  stochastic growth and division dynamics of cells at different temperatures can be obtained from the dynamics at any other temperature by scaling a single characteristic timescale. Panels (a, b) are taken from~\cite{2014-iyer-biswas-dn}. Panel (c) is taken from~\cite{2018-jafarpour-ad}.}
    \label{fig:differenttemperatures}
\end{figure}

This scaling was demonstrated in \textit{C. crescentus} cells through high-precision SChemostat experiments at different temperatures, establishing that the division time distributions, which can vary substantially across physiologically accessible temperatures (Fig.~\ref{fig:differenttemperatures}a), nevertheless collapse to the same distribution when rescaled by their respective mean values (Fig.~\ref{fig:differenttemperatures}b)~\cite{2014-iyer-biswas-dn}! We label this ``mean-rescaling''. Since the population age distribution is determined by the dynamics of cell division, population age distributions across a range of temperatures also undergo mean-rescaling (Fig.~\ref{fig:differenttemperatures}c)~\cite{2018-jafarpour-ad}. Mean-rescaling of both size and division time distributions across different conditions have also been observed in \textit{E. coli}~\cite{kennard2016individuality,shimaya2021scale}, hinting at the universality of a characteristic timescale determining growth and division dynamics across different bacterial species. Scaling has also been shown for the distributions of protein copy numbers measured in populations under a wide range of conditions in \textit{E. coli} by first subtracting the mean and then dividing by the standard deviation~\cite{brenner2015single}.

Furthermore, by solving for the evolution of population distributions for asymmetrically dividing cells such as \textit{C. crescentus}, difficult-to-observe details about the swarmer phase can be extracted by combining population growth and SChemostat experiments~\cite{2018-jafarpour-ad}. Thus, the swarmer-to-stalked transition time in \emph{C. crescentus} was evaluated at different temperatures and shown to be controlled by the same emergent timescale in both stalked and swarmer cells~\cite{2018-jafarpour-ad}.

A practical consequence of these observations is that high-precision experiments need not be repeated at every growth condition to find distributions of cellular-level quantities such as division times. Single-cell distributions can be characterized by high precision experiments performed at a single growth condition. Then, simple and much less laborious population growth experiments can be performed at any additional growth condition of interest to measure the multiplicative change to the emergent time scale. The single cell distributions corresponding to this new condition can be obtained from the previously measured distributions (in the old growth condition) by simply stretching/compressing them by the multiplicative factor obtained from the population growth experiment. Alternately, oscillations in population numbers of an initially synchronized population of cells can be used to derive estimates of not only the division time scale, but also other parameters of the single-cell division time distribution~\cite{2018-jafarpour-ad}.

\subsubsection{Revisiting the constraining motif: The Stochastic Hinshelwood Cycle (SHC)}\label{sec:SHC2}
How can the SHC motif emerge from a confusing tangle of underlying biochemical networks whose details may vary not just from organism to organism, but also from condition to condition? To answer this question, we note that a large family of reaction networks with autocatalytic loops reach steady-state exponential growth. Furthermore, the exponential growth rate of such a generalized Stochastic Hinshelwood Cycle is set by the growth rate of the Hinshelwood Cycle corresponding to the strongest autocatalytic loop, i.e., the loop whose geometric mean of rates is the largest~\cite{2014-iyer-biswas-qy}. Thus, once the constraint motif of a Hinshelwood autocatalytic loop dominates in the cellular metabolism, other details of the metabolic network are deconstrained, i.e., they can vary over a large variety of possibilities from condition to condition or from organism to organism, and yet yield robust exponential cell size growth.

\section{INTERGENERATIONAL SCALING LAWS GOVERNING STOCHASTIC GROWTH AND DIVISION OF INDIVIDUAL CELLS}\label{sec:intergenerational}

We now turn our attention to intergenerational scaling laws, that is, quantitative relationships that hold across multiple generations of a single cell's history.

\subsection{Stochastic intergenerational homeostasis of an individual cell's size}

\subsubsection{Historical background} 

The outstanding challenge of understanding homeostasis in living systems in the face of system (internal) and environment (external) fluctuations is that this process is not only \emph{dynamic} but also inherently \emph{stochastic}. It is intuitively appealing that a cell's size must be homeostatic: static snapshots of bacterial populations yield cell size distributions that fall within some narrow range defined by the species and condition-specific typical size. Yet, as we detail below, extant perspectives that build on deterministic frameworks and then introduce ``noise added on top'' miss essential aspects of how homeostasis is actually achieved and maintained, as revealed by high-precision, long-term dynamical data. These details are essential to develop a precise framework describing stochastic intergenerational homeostasis.

\subsubsection{Experimental evidence for stochastic intergenerational homeostasis}
\begin{figure}[h]
    \centering
    \includegraphics[width=.65\textwidth]{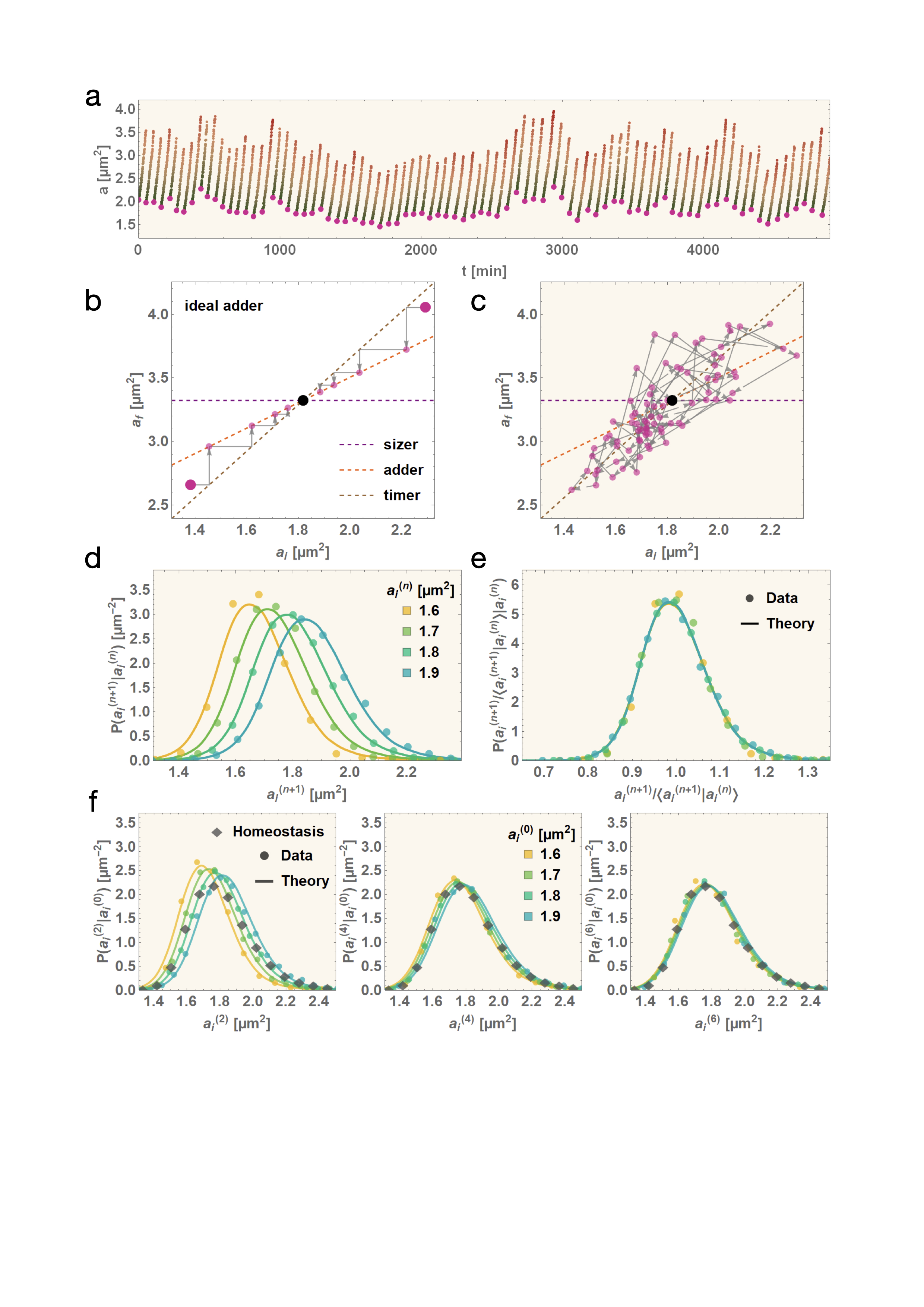}
    \caption{\textbf{Emergent simplicities in stochastic intergenerational homeostasis of cell sizes.} \textbf{(a)} A typical trajectory of cell size (cross-sectional area) for a \textit{C. crescentus} cell in complex media at 34$^\circ$ C. The initial sizes (sizes at birth) are indicated by maroon dots. \textbf{(b)} The intergenerational initial versus final size trajectory is plotted for two idealized ``theoretical'' cells  following the adder model (constant size added between divisions), showing the convergence to population mean. \textbf{(c)} The initial versus final size trajectory for the cell in (a). The traditional quasi-deterministic sizer--adder--timer paradigm of homeostasis in (b), marked as dashed lines, proves inadequate for capturing the dynamical process in a real cell. \textbf{(d)} The conditional distributions of the next generation's initial sizes, conditional on the current generation's initial sizes, are plotted for different values of the current generation's initial sizes. The points show the experimental measurements, while the lines show the predictions from the model combining the SHC with the thresholding of a protein to trigger constriction. \textbf{(e)} The distributions in (d), when rescaled by their mean values, collapse onto the same distribution. This scaling law determines the stochastic evolution of initial sizes across successive generations, leading to homeostasis. \textbf{(f)} For cells starting from four different initial sizes (represented by different colors), the distributions of initial sizes after 2, 4, and 6 generations are plotted from left to right. With increasing number of generations, these distributions all converge to the steady-state distribution of initial sizes (diamond markers) irrespective of the starting initial sizes. Data are taken from~\cite{2014-iyer-biswas-dn} and model predictions from~\cite{2023-architecture}.}
    \label{fig:homeostasis}
\end{figure}

It has only recently been rigorously established via high-precision experiments that cell sizes are maintained over many generations~\cite{2023-emergentsimplicity}. Since a cell's size changes over the course of a generation, a precise analysis requires the choice of a representative size within each generation, that can be compared to a comparable size in other generations. We choose this representative size to be the initial cell size of a given generation (Fig.~\ref{fig:homeostasis}a), and consider its intergenerational dynamics. Quasi-deterministic models of homeostasis within the sizer--adder--timer paradigm envision the dynamics of initial cell sizes as an orderly regression to the population mean (Fig.~\ref{fig:homeostasis}b), but this does not reflect the trajectories of actual cells measured from high-precision, long-term experiments using the SChemostat (Fig.~\ref{fig:homeostasis}c). To capture these stochastic dynamics, consider the conditional distributions of the next generation's initial sizes, conditional on the current generation's initial size (Fig.~\ref{fig:homeostasis}d). When the conditional distributions for different values of the current generation's initial size are rescaled by their respective mean values, they collapse onto a single distribution (Fig.~\ref{fig:homeostasis}e). In addition, the conditional distributions of the initial sizes after $n$ generations, conditional on the current generation's initial size, all converge to the steady-state distribution of initial sizes for $n \gg 1$ (Fig.~\ref{fig:homeostasis}f). Thus, not only does the population size distribution remain invariant in balanced growth conditions, but subpopulations characterized by different sizes in the starting (first-observed) generation all converge to the same homeostatic distribution of cell sizes after sufficiently many generations have elapsed. This nontrivial probabilistic ergodic behavior conclusively demonstrates stochastic intergenerational homeostasis of cell size~\cite{2023-emergentsimplicity}.

\subsubsection{Beyond the mythical average cell}
Before continuing to the emergent simplicities arising in stochastic intergenerational homeostasis of bacterial cells, we briefly discuss the historically popular sizer--adder--timer paradigm of cell size homeostasis~\cite{spiesser2012size,deforet2015cell,taheri2015cell,sauls2016adder,logsdon2017parallel,willis2017}, which was formulated following older studies of cell size control~\cite{1981-nurse,1987-tyson,2000-nurse}. These terms refer to purported decision-making strategies of cells based on which they undergo division: sizer cells measure cell size and commit to division when a critical preset size is reached, adder cells also measure cell size but commit to division when a critical preset size has been added, and timer cells measure the time elapsed since the last division and divide when a critical preset time has elapsed. While satisfyingly intuitive, this anthropomorphic view of cellular decision-making is difficult to verify through experiments probing underlying mechanisms, and furthermore fails to capture basic phenomenology exhibited by high-precision experimental data. Importantly, the sizer--adder--timer perspective on homeostasis does not accommodate stochasticity in a principled way.To illustrate, we discuss two central predictions of this framework. First, the mean final pre-division cell size versus the corresponding (same generation) initial cell size is predicted to be a straight line, with three discrete allowed values for the slope: $0$ for the sizer, $1$ for the adder, and $1/r$ for the timer, where $r$ is the division ratio. Experimental data do show a linear trend over the measurable dynamic range; however, the slopes extracted, for example for \textit{C. crescentus} under different growth conditions, do not match any of the three canonical predicted values~\cite{jun2015cell}. Moreover, they are found to vary, apparently without restrictions on continuity, from organism to organism and even from condition to condition for the same organism~\cite{jun2015cell,2023-emergentsimplicity}. In an attempt to accommodate these experimental observations while retaining the sizer--timer--adder paradigm, mixer models have been proposed~\cite{Modi2017}. Second, a cell whose size is not equal to the target (homeostatic) size in a given generation``heals'' to the target size in a deterministic manner over successive generations for sizer and adder cells: the sizer heals in a single generation while the adder heals exponentially with the deviation from the target size decreasing by a factor $r$ over each successive generation~\cite{amir2014cell,jun2015cell,lin2017effects}. This prediction is not borne out by high-precision data of individual cell size trajectories, which instead reveal a stochastic ergodic exploration of the population-level homeostatic cell size distribution~\cite{2023-emergentsimplicity}. These points motivate the need to revisit and revise the epistemology of homeostasis to incorporate the observed emergent simplicities in intergenerational individual cell trajectories and develop complementary conceptual frameworks.
 
\subsubsection{Intergenerational cell size dynamics are Markovian}\label{sec:MarkovSize}
One of the earliest models of growth and division assumed a lack of correlation between cell sizes at subsequent division events~\cite{Koch1962}, but this remained an unproven hypothesis until the availability of high-throughput, high-precision single-cell data, which established the stronger conclusion that cell size evolution is Markovian under balanced growth conditions: the next generation's conditional initial size distribution, conditioned on the current generation's initial size, is independent of the size from any previous generation~\cite{2023-emergentsimplicity}. This behavior, although simple, provides an important piece of an interesting puzzle: do bacterial cells use the memory of past events to modify behavior, and if yes, then how? Indeed, since the initial sizes (sizes at birth) are observed to follow Markovian dynamics under constant balanced growth conditions, we conclude that cell size homeostasis is maintained through \emph{elastic} adaptation~\cite{2023-emergentsimplicity}. In contrast, the individual cell's growth rate undergoes non-Markovian intergenerational dynamics and is maintained through \emph{plastic} adaptation~\cite{2023-nonmarkov} (see Section~\ref{sec:NonMarkovian}).

\subsubsection{Intergenerational scaling governs stochastic cell size homeostasis}\label{sec:InterScalingLaw}
When the conditional distribution of \emph{final} cell sizes, conditioned upon a given value of the initial (post-division) cell size, is rescaled by its initial size-dependent mean, the resulting rescaled distribution is independent of the specific value of the initial size used~\cite{2023-emergentsimplicity}. Since the division ratio is not significantly correlated with cell size, this translates to the following rule for intergenerational evolution of cell size: the mean-rescaled conditional distribution of \emph{initial} cell sizes, conditioned on the initial cell size of the previous generation, is also independent of the specific initial size of the last generation. Given this mean-rescaled distribution, the only information required to extract the original conditional distributions of the final size (or the next generation's initial size) is the stretching factor, which gives the functional dependence of the mean final size (or the next generation's mean initial size) on the initial size. As previously noted, this mean calibration curve is experimentally found to be a straight line over the measurable range of sizes, with a slope that is generally different from the discrete options corresponding to the sizer--adder--timer paradigms. Remarkably, following the discovery of this intergenerational law using data for \textit{C. crescentus} cells in the high-precision SChemostat setup~\cite{2023-emergentsimplicity} and validation in the Mother Machine~\cite{ziegler2023scaling}, our reanalysis of extant published data on \textit{E. coli} and \textit{B. subtilis} show that these organisms obey the same intergenerational scaling law~\cite{2023-precisionkinematics}.

\subsubsection{Precision kinematics and breakdown of homeostasis}\label{sec:precisionkinematics}
Denoting by $a_{n}$ the stochastic variable corresponding to initial cell size in the $n^{\text{th}}$ generation, the intergenerational emergent simplicities, combined with the linear character of the calibration curve connecting the initial cell size to the mean initial cell size in the next generation, yield the following stochastic map~\cite{2023-precisionkinematics,2023-emergentsimplicity}
\begin{equation}\label{eq-InterEvolution}
a_{n+1} = s_{n}\le(\a a_{n} + \b\ri),
\end{equation}
where $\a$ and $\b$ are experimentally measured numbers characterizing the linear calibration curve, and $s_{n}$ is a random variable whose probability distribution, denoted by  $\Pi(s)$, is the experimentally measured mean-rescaled conditional initial cell size distribution conditioned on the initial size of the previous generation (independent of $a_{n}$, see Section~\ref{sec:InterScalingLaw}). The distinct random variables denoted by $s_{n}$ for different values of $n$ are independent of each other, while being drawn from the same distribution, $\Pi(s)$. High-precision SChemostat data show that Eq.~1 indeed correctly describes the intergenerational evolution of the distributions of cell sizes $a_{1}, a_{2}, \ldots $ given an initial starting size value $a_{0}$~\cite{2023-emergentsimplicity}. This was also confirmed by Mother Machine data, for the rod-shaped mutant $\Delta$creS strain of \textit{C. crescentus}~\cite{ziegler2023scaling} as well as for \textit{E. coli} and \textit{B. subtilis}~\cite{2023-precisionkinematics}.

Let us represent the $k^{\text{th}}$ moment of $\Pi(s)$ by $m_{k}$ and the minimum value of $s$ beyond which $\Pi(s)$ is always zero by $s_{\text{max}}$. By definition, $m_{0}=m_{1}=1$. The moments $\le\{m_{k}\ri\}$ characterize the strength of noise (stochasticity) against which homeostatic maintenance of cell size is achieved through Eq.~1. Mathematically, cell size homeostasis is ensured if, for an arbitrary cell size value $a_{0}$ at the initial generation, the distributions of $a_{1}, a_{2}, \ldots $ successively evolve towards an $a_{0}$-independent, well-behaved homeostatic distribution (with zero probability of finding arbitrarily large yet otherwise normal cells). Equivalently, for cell size homeostasis to occur, all moments of cell size asymptotically tend to finite values independent of $a_{0}$, i.e., $\lim_{n\to\infty}\le\langle (a_{n})^{k} \ri\rangle$ is finite and $a_{0}$-independent for all $k = 1, 2, \ldots$.

These moment-based homeostasis conditions yield remarkably simple criteria for homeostasis in terms of the experimentally observed slope, $\a$, and noise strengths $\le\{m_{k}\ri\}$: for the $r^{\text{th}}$ moment of the cell size to converge to an $a_{0}$-independent finite value, we must have $|\a|<\le(m_{k}\ri)^{1/k}$ for $k=1,2,\ldots r$ ~\cite{2023-precisionkinematics}. Furthermore, the maintenance of cell size homeostasis, requiring the $a_{0}$-independent convergence of all moments of cell size upon successive applications of Eq.~1, necessitates an even simpler criterion: $\a < 1/s_{\text{max}}$ (or $s_{\text{max}} < 1/\a$)~\cite{2023-precisionkinematics}. This homeostasis condition is equivalent to requiring that the noise distribution lie entirely to the left of $1/\a$ on the plot of $\Pi(s)$ vs.\ $s$. This condition is indeed found to be true for extant experimental data~\cite{2023-precisionkinematics}.

Finally, we note that experimental data show that $\Pi(s)$ is typically remarkably close to saturating the bound of $1/\a$ on the domain where it is nonzero~\cite{2023-precisionkinematics}. While this intriguing observation of quasi-perfectly tuned homeostasis merits rigorous analysis in future studies, here we speculate that it may well represent a trade-off involving the cost of maintaining a small rigid value of $\a$ against the cost of violating size homeostasis. Perhaps it is indicative of an organismal strategy of maintaining a sliver of the population of cells at the exploratory frontier, to undertake high-risk, high-reward explorations for possible evolutionary advantage~\cite{DeGroot2023}.

\subsubsection{Constraining motif: the $Q$--$X$ module}\label{sec:QX}
The intergenerational emergent simplicities described above are captured by a minimal analytically solvable mechanistic model~\cite{2023-architecture}. The model may be motivated as follows. Direct cell size control can be na{\"i}vely implemented by thresholding the cell size, a problem whose stochastic implementation as a first passage problem is analytically tractable, but yields predictions for division time distributions whose tails do not quite agree with experimental data~\cite{2014-iyer-biswas-qy,2014-iyer-biswas-dn,2016-iyer-biswas-rz}. This is not surprising since, presumably, it is costly to maintain machinery to directly measure and threshold cell size; thus, such tight control may not be cost effective. Loosening this tight control somewhat (and thus diminishing costs) can serve as motivation to instead use a measurable reporter, $Q$, whose copy numbers report cell size by requiring that it grow at a rate proportional to the cell volume. The division process is triggered when $Q$ hits some threshold $\Theta$, subsequently taking some random time $T$ to complete. Denoting by $X$ the effective Stochastic Hinshelwood Cycle variable proportional to cell size, the growth dynamics of this minimal setup are characterized by the stochastic processes
\begin{equation}
X \stackrel{k_{X}}{\to} X+X, \quad X \stackrel{k_{Q}}{\to} X + Q,
\end{equation}
occurring since the start of the cell cycle until $Q$ hits a threshold $\Theta$. Following this thresholding event, for a stochastic time period $T$, growth continues according to the same scheme but with a different rate $k_{Q}'$ associated with the production of $Q$. Finally, the cell divides and partitions $Q$ and $X$ between the daughter cells.

Using analytic methods, including a newly developed technique based on a stochastic rescaling of time, this model yields~\cite{2023-architecture}: 
\begin{enumerate}[label=(\alph*)]
\item The mean conditional post-division initial size in the next generation, conditioned on the current generation's initial cell size, is a linear function of the latter. The corresponding slope, equal to $\a$ in Eq.~1, is a continuously variable quantity that depends on the specific parameters of the model and is thus \emph{not} a reflection of the strategy of cell division.
\item The sequence of cell sizes attains homeostasis despite stochasticity associated with intragenerational growth and with intergenerational variation of parameters in the model, as long as the noise strength is not excessive. In analogy with the experimentally relevant conditions summarized in Section~(\ref{sec:precisionkinematics}), whose violation lead to a breakdown of homeostasis, straightforward conditions on the stochastic parameters were analytically derived for maintenance of homeostasis of each moment of the size distribution~\cite{2023-architecture}. These conditions hold, given a value of $\a$ such that $|\a|<1$, until noise in parameters exceed some bounds.
\end{enumerate}
Thus, this minimal model exhibits stochastic intergenerational cell size homeostasis, with breakdown of homeostasis occurring when noise crosses a certain threshold. Furthermore, when the numbers of $Q$ and $X$ are not small, when the numbers of $Q$ are much less than those of $X$ (recall that the motivation for introducing $Q$ is to mitigate the cost of thresholding size directly) and for an appropriate division rule~\cite{2023-architecture}:
\begin{enumerate}[resume, label=(\alph*)]
\item The intergenerational scaling law (Section~(\ref{sec:InterScalingLaw})) emerges from this minimal model.
\item The conditions for avoiding breakdown of homeostasis reduce to those derived in Section~(\ref{sec:precisionkinematics}).
\item The conditional division time distribution, conditioned on the initial size, and also the steady-state division time distribution predicted by this model exactly match the corresponding experimental distribution, solving an outstanding puzzle that arose from mismatch between data and a size-thresholded division model of stochastic exponential growth~\cite{2014-iyer-biswas-qy,2014-iyer-biswas-dn}!
\end{enumerate}

The $Q$--$X$ ``plug-and-play'' motif is thus a compelling candidate for defining the minimal architecture that ensures stochastic intergenerational homeostasis. While this motif constraint represents an indispensable attribute of the core cell size regulation machinery, other details of cellular regulation and system-specific instantiations of the model are ``deconstrained'' within the limits defined by the aforementioned conditions of breakdown of homeostasis~\cite{2023-precisionkinematics,2023-architecture}, while still leaving intact a properly functioning cell size homeostasis machinery. A prime candidate for $Q$ is the highly conserved protein FtsZ~\cite{2023-architecture}, which initiates and drives constriction in bacterial cells~\cite{Aaron2007,Adams2009,Goley2011}; its characteristics are consistent with predictions for cell size at the First Passage Time of FtsZ copy numbers to cross a fixed threshold, followed by a roughly constant amount of time between onset of constriction until cell division~\cite{govers2024apparent}. However, further high-throughput measurements linking the dynamics of FtsZ, cell size, and cell growth are needed to confirm this hypothesis.

\subsubsection{Generalization across conditions and experimental realizations}
We close this discussion by considering how high-precision data from one experimental design translates to another, in a principled manner. A side-by-side comparison of the same strain of \textit{C. crescentus} growing in both the SChemostat and Mother Machine devices was performed under balanced growth conditions to generate a dataset suitable to explore this question~\cite{ziegler2023scaling}. Although the distributions of division times, the division ratio, and absolute sizes were affected by mechanical confinement in the Mother Machine, the growth rate distribution remained invariant. Yet remarkably, the deviations in growth rate precisely balanced out the deviations in division time, resulting in the combination $r e^{\kappa \tau}$ (which gives the fold change in cell size between consecutive births) having the same population-wide distribution in both the SChemostat and the Mother Machine, with $r$, $\kappa$ and $\tau$ denoting the division ratio, exponential growth rate and division time, respectively. This, combined with the observed mean rescaling of initial size distributions across the experimental conditions~\cite{ziegler2023scaling}, results in the remarkable reproduction of the \emph{same} emergent interdivision scaling law described in Section~(\ref{sec:InterScalingLaw}) (but applied to mean-rescaled initial sizes instead), and with the same mean-rescaled homeostatic size distribution.

\subsection{Non-Markovian memory and plastic adaptation of the single cell growth rate}\label{sec:NonMarkovian}
A feature of the SChemostat technology is that it enables the study of statistically identical non-interacting cells responding to precisely defined \emph{time}-varying growth conditions. A preliminary understanding of how \textit{C. crescentus} cells respond to time-varying growth conditions is presented below~\cite{2023-nonmarkov,2023-timedependent}. Before proceeding to time-dependent environmental changes, however, we briefly consider the consequences of one of the intergenerational emergent simplicities: cell size evolves through a Markov process (see Section~(\ref{sec:MarkovSize})). This was experimentally established by demonstrating that the measured conditional probabilities $P(a_{n+1}|a_{n}, a_{n-k})$ for $k \geq 1$ are independent of $a_{n-k}$~\cite{2023-emergentsimplicity}. In a minimal model of growth and division that involves only the cell size, \emph{behavior} is encoded in the functional form of the Markov conditional probability $P(a_{n+1}|a_{n})$. Thus, due to the Markovian nature of cell size, a change of growth conditions leads to a change in behavior, and restoring growth conditions restores behavior. Adaptation---the change of behavior---involves no memory of the past in this scenario. This is known as elastic adaptation.

A more interesting scenario would involve an organism-level observable whose behavior depends on the intergenerational history of the cell, i.e., it displays \emph{plastic} adaptation. An indication of such behavior is non-Markovian evolution. Statistical analysis of high-precision SChemostat data indicate that the instantaneous exponential growth rate displays such non-Markovian behavior. Denoting by $\k_{n}$ the exponential growth rate in the $n^{\text{th}}$ generation, the conditional probabilities $P(\k_{n+1}|\k_{n}, \k_{n-k})$ (with $k \geq 1$) were found to exhibit a measurable dependence on $\k_{n-k}$ for $k$ on intergenerational timescales~\cite{2023-nonmarkov}. In other words, complex memoryful adaptive behaviors on intergenerational timescales can now be observed and characterized in individual bacterial cells under different growth conditions.

\subsubsection{Beyond time-invariant growth conditions}

Aspects of plastic response to time-varying conditions have been characterized and reported in \cite{2023-nonmarkov}, wherein \textit{C. crescentus} cells inside the SChemostat were subjected to an abrupt change to a growth condition never experienced by the cells throughout their complete histories. Two observations from this work are noteworthy. First, a single time-dependent timescale, proportional to the inverse instantaneous mean growth rate across the population, controlled growth and division dynamics across many tens of generations of slow adaptation after balanced growth was disrupted by the switch. Once the measured strongly time-dependent population age distributions were appropriately rescaled using the observed time-varying mean instantaneous growth rate~\cite{2023-timedependent}, they became time-invariant as if the cells had always been in balanced growth. This is a significant extension of scope of the intragenerational emergent simplicity discussed in Section~(\ref{sec:part1intra}). Much of the added complexity in a living organism responding to time-varying environmental changes is encoded in a single composite emergent degree of freedom! Second, memory of the switch was encoded in the binary response of individual instantaneous cell growth rates. Slow \emph{resetting} of memory occurred as individual cells unidirectionally and stochastically migrated from one branch of response corresponding to a lower growth rate to another corresponding to a higher growth rate.

Recall that, in the SHC model, the exponential growth rate is determined by the geometric means of the reaction rates of the dominant autocatalytic cycle (which themselves can stochastically vary from cell cycle to cell cycle). An abrupt shift in nutrient conditions is likely to change the dominant cycle, as the cell adjusts its metabolic network to optimize the suite of enzyme pathways used to process available nutrient sources, a phenomenon with well-characterized effects on population growth rate~\cite{Monod1942,Kjeldgaard1958}. In the case of a nutrient downshift (from nutrient-rich to -poor media), instantaneous growth rate has been observed to instantaneously fall to zero, before eventually recovering to the new steady state~\cite{2023-nonmarkov,erickson2017global}. In \textit{E. coli}, the reason for this suboptimal recovery appears to be the rigid strategy of protein synthesis allocation~\cite{erickson2017global}; this is consistent with the notion a switch in the dominant cycle to process a new nutrient profile. The initial fall to zero would thus mark the loss of the previous dominant cycle into a transient state. The observed unidirectional stochastic migration from the branch of response corresponding to a lower growth rate to the one corresponding to a higher growth rate could correspond to the establishment of the new dominant cycle. Since the constituent reactions in the new cycle would differ, memory encoded in the rates of the previous dominant cycle would no longer affect the growth rate, effectively resetting the memory in the overall growth rate. The timescale of recovery to balanced growth could shed light on the complexity of the underlying SHC, since it has been shown that this recovery time is longer the more complex the  interdependence of the various processes in the SHC~\cite{1952-hinshelwood}. Recent experiments of \textit{E. coli} exposed to a step-like change in nutrient profile have observed dynamic fluctuations of a key metabolite that propagate to other cellular processes~\cite{Bi2023}; performing such experiments alongside high-precision growth studies will yield crucial connections to biological mechanisms and shed light on the identities of the biochemical reactions constituting the dominant cycle.

\section{CONCLUSION AND FUTURE PERSPECTIVE}

General principles of biology tend to be rare, owing to the idiosyncratic life histories of different species. An obvious exception is evolution by drift and selection, which provides a powerful framework encompassing the full diversity of life. Allometry describes systematic and regular scaling of organismal form and function corresponding to variations in body size, which arises from the connection between strong physical constraints and optimal physiology~\cite{West2017}. These interspecies ``scaling'' relationships represent a certain type of universality, based on shared dominant constraints. However, one of the challenges to a general theory of living systems has been that the sources of cross-species regularity tend to be statistical phenomenologies, rather than being derived from first principles. Bridging constitutive dynamics with macroscale regularity requires an approach that is both sensitive to genetic circuitry and also mappable onto low-dimensional, organism-level ``state variables'', which provide the most efficient, and perhaps most predictive, descriptions of biology.

In this review, we have illustrated how an integrated approach uniting high-precision measurements, data-informed insights, and physics theory can successfully address outstanding questions of fundamental biology. Focusing on the phenomenon of stochastic growth and division processes in the simplest living organism (the bacterial cell), we have walked through a procedure for analyzing high-throughput, high-precision dynamic datasets to identify emergent simplicities, in particular various scaling laws, that provide new insights into a long-standing problem (that of cell size homeostasis). These scaling laws include the mean-rescaled cell size distributions across cell ages, the mean-rescaled distributions of division time and cell age across growth conditions, and the mean-rescaled distributions linking division ratio to both growth rate and division time ($r e^{k \tau}$) across experimental modalities. Recasting the question from a stochastic, intergenerational viewpoint (i.e., one that considers the entire life histories of individual cells without recourse to \textit{a priori} mechanistic assumptions), and taking advantage of identified emergent simplicities to achieve dimensional reduction of the problem, permits a reformulation that captures the inherent stochasticity of individual cells.

Identification of discrete modes by which homeostasis is maintained---in particular, via reflexive (elastic) adaptation of cell size and reflective (plastic) adaptation of growth rate---provides important insights into key system constraints that govern living bacterial cells, with additional implications for the design of functional adaptive synthetic homeostats. The observation of non-Markovian dynamics in single-cell growth rates implies the existence of intergenerational memory and plastic adaptation in these simple organisms, presenting an exciting opportunity to uncover the mechanistic basis of learned behavior in a single-celled life form without neurons~\cite{Rajan2023,2023-wright-rc}. In particular, the generalization from steady state to complex, time-varying growth conditions presents many promising avenues for future research. This includes transitions between different growth regimes, growth under stressful conditions that lead to cell death and aging (such as starvation or exposure to antibiotics), and growth in environments that permit non-trivial collective behaviors emerging from interactions between cells.

Perhaps the most striking finding is the emergence of a single cellular unit of time that dictates the stochastic growth and division dynamics of individual cells. With the appreciation that each individual bacterium is playing out its own unique and whimsical rhythm, by observing a large number of these living timekeepers we start to discern an evocative polyphony, an emergent song that reflects the underlying simplicities.

The confluence of precision microfluidics, quantitative long-term, live-cell imaging, and bespoke data-analysis pipelines now enable a level of quantitative rigor far surpassing that available to last century's pioneers in quantitative growth studies. It is up to us to not lose sight of the forest for the trees; taking the physics approach of identifying unifying themes in apparently disparate phenomena continues to be a fruitful approach.

\section*{DISCLOSURE STATEMENT}
The authors are not aware of any affiliations, memberships, funding, or financial holdings that might be perceived as affecting the objectivity of this review.

\section*{ACKNOWLEDGMENTS}
We thank Purdue University Startup funds, Purdue Research Foundation, the Purdue College of Science Dean's Special Fund, and the Showalter Trust for financial support. K.J., and S.I.-B. acknowledge support from the Ross-Lynn Fellowship award. K.J. and S.I.-B. acknowledge support from the Bilsland Dissertation Fellowship award. S.I.-B. thanks the Harvard Medical School's Department of Systems Biology and Jeremy Gunawardena for graciously hosting her as an extended visitor during early stages of this work.

%
\renewcommand{\refname}{LITERATURE\ CITED}

\pagebreak

\end{document}